# Kagome silicene: a novel exotic form of two-dimensional epitaxial silicon

Y. Sassa[1,2]*, F.O.L. Johansson[2], A. Lindblad[2], M.G. Yazdi[3], K. Simonov[2], J. Weissenrieder[3], M. Muntwiler[4], F. Iyikanat[5], H. Sahin[6]*, T. Angot[7], E. Salomon[7]*, and G. Le Lay[7]

[1] Department of Physics, Chalmers University of Technology, SE-412 96 Göteborg, Sweden

[2] Department of Physics and Astronomy, Uppsala University, SE-752 36 Uppsala, Sweden

[3] Materials and Nanophysics, KTH Royal Institute of Technology, Electrum 229, SE-16440 Kista, Sweden

[4]Paul Scherrer Institute, Swiss Light Source, 5232 Villigen PSI, Switzerland

[5]Department of Physics, Izmir Institute of Technology, 35430 Izmir, Turkey

[6]Department of Photonics, Izmir Institute of Technology, 35430, Izmir Turkey

[7]Aix-Marseille Univ., CNRS, PIIM, Marseille, France

*Corresponding authors: yasmine.sassa@chalmers.se; hasansahin@iyte.edu.tr; eric.salomon@univ-amu.fr

**Abstract**

Since the discovery of graphene, intensive efforts have been made in search of novel two-dimensional (2D) materials. Decreasing the materials dimensionality to their ultimate thinness is a promising route to unveil new physical phenomena, and potentially improve the performance of devices. Among recent 2D materials, analogs of graphene, the group IV elements have attracted much attention for their unexpected and tunable physical properties. Depending on the growth conditions and substrates, several structures of silicene, germanene, and stanene can be formed. Here, we report the synthesis of a Kagome lattice of silicene on aluminum (111) substrates. We provide evidence of such an exotic 2D Si allotrope through scanning tunneling microscopy (STM) observations, high-resolution core-level (CL) and angle-resolved photoelectron spectroscopy (ARPES) measurements, along with Density Functional Theory calculations.

The two-dimensional Kagome lattice, which consists in corner-sharing rigid triangles, possesses fascinating magnetic properties and an exotic correlated electronic structure, possibly hosting Dirac electronic states [1, 2], that could lead to a topological insulator [3]. On-surface Kagome lattices are often engineered in 2D metal-organic systems [4]; they might be tailored also using a molecular pattern [5]. Rare examples concern experimentally realized inorganic networks [6]. 2D Kagome lattices for phosphorous and for group-IV elements have been predicted from first-principles calculations [7,8], but not achieved experimentally; it is worth mentioning a purely electronic Kagome lattice observed on a twisted multilayered silicene surface [9] .In this work, we have formed directly at room temperature an epitaxial silicon epitaxial sheet in a Kagome lattice on the Al(111) substrate.

In the last years, graphene-like forms of Si, Ge and Sn, coined silicene, germanene, and stanene, have been successfully grown, mostly on the (111) surfaces of silver and gold single crystals [10,11,12,13]. A buckled 2×2 reconstructed germanene layer, matching a 3×3 superstructure of an aluminum (111) substrate, was first obtained by Derivaz *et al.* in 2015 [14]. Years ago, a 3×3 Al(111) superstructure was initially observed in Low-Energy Electron Diffraction (LEED) patterns upon depositing Si onto a clean Al(111) surface [15], but since then, no new experimental result on this system has been published. Recent studies based on molecular dynamics calculations [16] have demonstrated that a 4×4 honeycomb, or polygonal, silicene monolayer similar to that obtained on Ag(111) can be stably formed on the Al(111). Nevertheless, in the present work, no such 4×4 superstructure was observed upon Si deposition, but a 3×3 phase, which is very intriguing and interesting, possibly featuring a novel 2D silicon phase. By analogy with the Ge deposition case, the observed Al(111)-3×3 reconstruction could suggest the formation of a 2×2 silicene sheet [14].

## Materials and Methods

**Sample preparation:** A monolayer of silicon was in situ deposited on a clean Al(111) crystal (Surface Preparation Laboratory, the Netherlands) at room temperature (RT) for eight minutes under a vacuum of $10^{-9}$ mbar (by sublimation of a piece of Si wafer through direct current heating). The clean Al (111) surface was obtained after several cycles of $Ar^+$ ion sputtering followed by annealing at 710 K. Immediately after deposition at RT, without further annealing, the surface quality was checked by LEED; then the sample was transferred under UHV to the XPS/ARPES and STM chambers.

**CL- and AR-PES experiments:** The CL- and AR-PES experiments were carried out at the PEARL beamline at the Swiss Light Source. The Si and Al 2p core levels were measured with a photon energy of 136 eV in normal emission and grazing emission. The ARPES spectrum was measured with a He lamp with a photon energy off 21.2 eV. Both CL- and AR-PES spectra were recorded using a SCIENTA R4000 electron analyzer with vertical slit setting. All data were recorded at room temperature.

**STM experiment:** The STM experiment were carried were carried out at the PEARL beamline at the

Swiss Light Source on the same samples used for the ARPES measurements. All the STM images were acquired at low-temperature (T= 70 K) using a standard SCIENTA-OMICRON LT-STM set-up. Various current and voltages were used and are specified in the figures captions.

**Computational Methodology**

The first-principles calculations were carried out using density functional theory (DFT) and the projector-augmented wave (PAW) method [17,18], as implemented in the Vienna ab-initio Simulation Package (VASP) [19,20]. In order to simulate the structure, a Kagome silicon sheet was placed on top of a 3×3 Al(111) surface. The Al(111) surface was modeled by five atomic layers. While the atoms of the Kagome silicon layer and the three upper Al layers were free to move, the atoms of the two bottom Al layer were fixed. The exchange-correlation energy was described by the generalized gradient approximation (GGA) using the Perdew-Burke-Ernzerhof (PBE) functional [21]. The van der Waals (vdW) correction to the GGA functional was included by using the DFT-D2 method of Grimme [22]. A plane-wave basis set with energy cutoff of 400 eV was used. Γ-centered k-point mesh was employed with grid sizes of $3 \times 3 \times 1$ and $9 \times 9 \times 1$ for the structural relaxation and the density of states (DOS) calculation, respectively. The total energy was minimized until the energy variation in sequential steps became less than $10^{-5}$ eV in the structural relaxation. The total force on each unit cell was reduced to a value of less than $10^{-4}$ eV/Å. At least 12 Å of vacuum space was added along the z-direction to avoid interactions between the adjacent cells. Analysis of the charge transfers in the structure was determined by the Bader technique [23].

**Results and Discussion**

In our experiments, we observed a sharp 3×3 LEED pattern (Fig. 1a) upon deposition of about one monolayer of Si onto a clean Al(111) surface kept at room temperature without any further annealing. The overview of the core-level spectra acquired from the clean Al(111) surface confirmed the absence of any signals from carbon and oxygen, the main contaminants of highly reactive aluminum. An analysis of the Al(111)3×3-Si surface by STM displays wide terraces (several tens of nanometers), separated by atomic steps. As can be seen from Fig. 1b, an ordered layer of Si covers most of the surface. Interestingly a height profile recorded in a region, where bare Al(111) and the Si layer coexist (Fig. 1c and 1d), indicates that the overlayer is quite smooth, exhibiting a corrugation of about 30 pm, and at about the same apparent height as the bare Al(111). This points to a mechanism where the top layer substrate atoms are initially displaced, a case already encountered upon the formation of a silicene adlayer on the Ag(111) surface [24]. The STM image of the Al(111)3×3-Si exhibits bright protrusions in a hexagonal arrangement. The distance between two neighboring protrusions corresponds to 8.6 Å, which is about three times the surface lattice parameters of Al(111), in accordance with the observed 3×3 LEED pattern (Fig. 1a). The unit cell of the 3×3 superstructure is represented by the black rhombus in fig. 1c.

Figure 2 presents two high-resolution STM images of the Al(111)3×3-Si surface. The bright protrusions are still dominating, but perusal reveals less bright features and dark holes in a rhombic 3×3 supercell (black rhombus). This indicates that the observed 3×3 superstructure is made of Si atoms at different heights, as in the case of the Al(111)3×3-Ge system [14,25,26]. Nevertheless, even though the system Si on Al(111) exhibits the same 3×3 surface periodicity with respect to Al(111) substrate, the STM images do not resemble those obtained for the Ge on Al(111) system. As a matter of fact, they do not exhibit a clear honeycomb pattern (corresponding to a 2×2 germanene sheet with two protruding Ge atoms matching the Al(111)3×3 supercell [14]), nor just a simple hexagonal pattern with only one protruding Ge atom, with a frequent tip-induced switch between the two patterns [26]. They also differ from the hexagonal structure obtained by Endo *et al.*, which corresponds to a √3×√3 germanene sheet matching a different Al(111)√7×√7 crystallographic supercell [27]. Thus, despite the same sized Al(111)-3×3 surface unit cell, this points to another atomic arrangement within the cell, different from a simple honeycomb array, when depositing silicon instead of germanium onto Al(111). From these STM images, a possible arrangement in a Kagome-like lattice can be suggested as depicted by the orange mesh in figure 2. Indeed, as described later in the manuscript, the observed superstructure which adopts a 3×3 surface reconstruction with respect to the Al(111), matches a (√3×√3)R30° reconstruction with respect to a pristine 2D Kagome lattice.

Core-level spectroscopy is generally very useful to further analyze such surface layers. Hence, we have performed high-resolution synchrotron radiation measurements of the Al 2p and Si 2p core levels photoelectron spectra at the PEARL beamline of the Swiss Light Source [28]. First of all, due to the small surface corrugation revealed by the STM studies and the small height difference between bare Al(111) and Si/Al(111), we have recorded both Al 2p and Si 2p spectra at normal and grazing emission. As depicted in fig. 3a and 3b, in the case of the Al(111)3×3-Si surface, the signal from the Si 2p doubles, with respect to the signal from the Al 2p, when changing the sample orientation from normal emission (NE) to grazing emission (GE) (60$^o$ off-normal), while the overall line shape of the Si 2p does not change much. Together with a significant attenuation of the Al 2p signal (fig. 3c) it confirms that Si atoms form an overlayer on top of the Al(111) surface, as in case of Ge deposition. [14,25,26]). The Al 2p spectrum from the clean bare Al(111) surface acquired in highly surface sensitive conditions (136 eV photon energy) appears nearly symmetric. After Si deposition, a shoulder appeared on the high-binding energy (BE) side of the Al 2p peak, due to interaction with the Si layer. More interesting is the Si 2p signal, which display five well resolved main peaks. Those features, which are remarkably sharp, reveal a well ordered and well defined atomic structure of the Si layer. It is however beyond the scope of this paper to give a precise assignment to each component. Nevertheless, the complex shape of the Si 2p signal suggests a complex atomic structure probably due to an intricate buckling of the Si atoms and their possible interaction with the Al substrate. However, from fig. 3d, one can note a marked increase of the relative weight of the lowest BE component when comparing spectra taken from normal to grazing emission. Since only this component exhibits such a behavior, it can be assigned to the bright protrusion seen at each Al(111)-3×3 supercell in STM imaging, associated to the Si adatom at the top of the dumbbell moiety (as discussed below).

To search for the detailed atomic geometry, we used DFT calculations (see methods for the computational details). The energetics and the simulated STM images of several initial configurations for the Si adlayer with hexagonal or trigonal symmetries were compared [7,29]. Among a wide variety of possible starting configurations, the STM data analysis suggests a Kagome type configuration. Such atomic arrangement would, however, correspond to a Al(111)-√3×√3R(30°) reconstruction. As depicted in fig. 4a and 4b, to match the actual Al(111)-3×3 supercell, an additional Si atom forming a dumbbell (DB) moiety arranged on top of Si layer was introduced, as initially proposed by Cahangirov *et al.* for silicene on Ag(111)[30]. Recently a DB model (coined by authors a « dumbbell silicene ») was reported to be the ground-state of 2D silicon [31]. The introduction of this DB was the key to obtain the most favorable structure in terms of cohesive energy calculations and STM simulations. Top and side views of the optimized geometric structure are shown in figs. 4a and 4b, respectively.

The structure, which we name Kagome silicene, is obtained when nine silicon atoms in a √3×√3 Kagome lattice are placed on top of an Al(111)-3×3 surface cell, and one additional Si atom, forming a DB moeity is introduced (hence, a total of 10 Si atoms). It corresponds to a total coverage ratio of 10/9 = 1.11, very close to that of the archetype 3×3 silicene phase matching the Ag(111)4×4-Si surface cell with 1.125 coverage ratio [10]. Within this structure, 2D Kagome silicene consists of four corner sharing triangles and one DB. While the triangles settle almost flat on the Al (111) surface, one silicon atom locates at the top level and forms a distorted tetrahedron with its three neighbors. The bond length between silicon atoms are in the range of 2.39-2.51 Å (for comparison the Si-Si distance in purely $sp^3$ hybridized bulk diamond type silicon is 2.35 Å), and the distance between the atoms of the dumbbell in the tetrahedron is 2.40 Å. The angles in the Kagome silicene sheet are between 56° and 141°, which points to mixed type hybridization, neither pure $sp^2$ (120° angle) nor pure $sp^3$ (109.5° angle). The height variation in the silicon layer is 1.13 Å. The closest Al-Si distance is 2.56 Å between the bottom dumbbell Si atom and a first layer Al atom at the interface is less than the sum of the atomic radii (2.605 Å), which points to a strong interaction between the sheet and the substrate.

Total energy calculations revealed that the cohesive energy of 2D Kagome silicene is 4.03 eV. The cohesive energy per atom was formulated as

$$E_{coh} = [n_{Al}E_{Al} + n_{Si}E_{Si} - E_{Al(111)+SL}]/N$$

where $E_{Al}$ and $E_{Si}$ denote the energies of single isolated Al and Si atoms, respectively. $E_{Al(111)+SL}$ represents the total energy of single-layer Kagome silicene and five-layer Al(111); $n_{Al}$ and $n_{Si}$ are the number of Al and Si atoms in the supercell, and N is the total number of atoms contained in the supercell. The simulated STM image representing the filled and empty states of the 2D Kagome silicene sheet are shown in Figs. 4(c) and 4(d), respectively. As seen in the figures there is a bright spot in the supercell due to the uppermost Si atom in the structure. These bright spots form the hexagonal array observed in the STM measurements. In accordance with the STM measurements, each supercell has large dark regions, which coincides with holes in the Kagome lattice. In addition, a careful analysis of the STM images depicted in Fig. 2 reveals on one part of the image (encircled region in the right-side image of fig. 2) that a bright protrusion is missing. Instead, one can only observe a dim protrusion in its place. This observation endorses the DB Kagome superstructure as this dim protrusion can be attributed to the lower Si atom usually located underneath a top Si atom.

For comparison with the core-level measurements, interatomic charge transfers in the structure were also calculated. Fig 5 (a) shows a side view of the 3D charge density map of the whole structure. It is seen that the charge is mainly distributed around and between the Si atoms. This indicates the covalent bonds between the Si atoms of the Kagome lattice. Moreover, charges around the bottom dumbbell Si atom tend to extend toward the intermediate region. In Fig. 5 (b) we show a side view of the 3D charge density differences, which are calculated by subtracting the charge of each individual Al and Si atoms from the final charge of the structure; the valence charges on the atoms are also shown. Bader charge analysis shows that charge transfer mainly occurs from Al(111) to the Kagome silicene layer. As seen from Fig. 5 (b), a large amount of charge accumulates in the intermediate region around the bottom dumbbell Si when the Kagome silicene sheet is placed on top of the Al(111) substrate, indicating a strong interaction. Despite the strong chemical bond that results from the high charge transfer around the bottom dumbbell Si atom between the silicene layer and Al(111), Kagome lattice with covalent bonds maintains a continuous layer form. The final charges on the interfacial Al layer are rather uniform (2.5 - 2.7 e), which complies with the single broad component of the Al 2p Cl, close to the bulk line (by just about 0.05 eV) on the high BE side. Instead, charges on the silicon atoms of the 2D Kagome silicene layer span a large range. While the Si adatom donates 0.2 e, its Si dumbbell partner receives 0.8 e. Therefore, final charges of the Si atoms are in the range of 3.8 - 4.8 e, which explains the 0.52 eV BE difference between the components 1 and 3 of the Si 2p CL.

Finally, Fig. 5 (c) shows the Brillouin Zone (BZ) of the 3×3 supercell structure (red hexagon) with respect to that of primitive Al(111)1×1 (blue hexagon). It is worth noting that $\bar{\Gamma}$ points of the (3×3) supercell coincide with $\bar{K}$ points of the primitive Al(111) BZ, while the $\bar{M}$ points of the two structures coincide. To investigate the electronic properties of the freestanding 2D

Kagome silicene sheet, the full band dispersions were also calculated (Fig. 5(d)). It is seen that several bands cross the Fermi level, demonstrating intrinsic metallic character of the 2D Kagome silicene on Al(111). The complete calculated band structure of Kagome silicene on Al(111)-3×3 (see Fig. 6( a)) yields a complex manifold, with no clear evidence of Dirac cones and flat bands. This makes the comparison with our set of experimental ARPES measurements taken with a He lamp (with finite energy and momentum resolutions at 21.2 eV), hardly tractable. However, we have measured nearly linear dispersive features, extending over 1 eV, at few $\bar{M}_{3\times3}$ points along $\bar{\Gamma}\bar{M}$ direction (see Fig. 6 (b)); they appear to correspond to crossings of the 3×3 folded band structure of bare Al(111) (Fig. 6 (c)). These features are actually similar to those observed in the case of monolayer silicon on Ag(111) by Y. Feng et al. [32]. We also note that for germanene on Al(111) the ARPES electronic structure measured recently reflected also just the subsurface Al layers [27]. Therefore, to further explore the detailed electronic structure of this 2D Kagome silicene sheet, additional ARPES data, acquired in more surface sensitive conditions with synchrotron radiation would be necessary.

To summarize, we have given compelling evidence of the epitaxial growth of Kagome silicene, a new artificial form of 2D silicon on an aluminum substrate. Due to strong interactions, the electronic properties differ from those expected for a free-standing sheet, but still may present promising characteristics, especially, most probably, after functionalization, e.g., by atomic hydrogen for instance. Indeed, for practical purposes like device fabrication, a transfer process could be necessary, and one fair advantage of synthesizing silicene on Al surfaces is that is can be much more cost-effective than growing it on Ir and Ag substrates, which led to the first realization of a Field Effect Transistor with an atom thin silicene channel operating at room temperature [33].

To conclude, in a daunting quest for novel 2D materials [34], the synthesis of a novel two-dimensional silicon allotrope opens highly exciting perspectives for new skeletal variants of silicene, germanene and stanene, possibly directly on semiconducting or insulating substrates.

**Acknowledgements**

GLL warmly thanks Prof. G. Margaritondo (Ecole Polytechnique Fédérale de Lausanne (EPFL), Switzerland), who observed for the first time by LEED in the late 80's the Al(111)3×3 reconstruction upon *in situ* Si deposition, for fruitful discussions, as well as Prof. H. Brune (EPFL), who showed unpublished STM images, acquired in the early 90's, and which stimulated undertaking a new study of the system. We are very grateful to L. Xian, and A. Rubio (Max Plank Inst. For the Structure and Dynamics of Matter, Hamburg, Germany), and S. Cahangirov (Bilkent Univ., Ankara, Turkey), who performed preliminary

DFT calculations, before passing the baton to FI and HS. We acknowledge the Paul Scherrer Institut, Villigen, Switzerland, for provision of synchrotron radiation beamtime at the PEARL beamline of the SLS. A.L. acknowledges the support from the Swedish Research Council (grant no. 2014-6463) and Marie Sklodowska Curie Actions (Cofund, Project INCA 600398). Y.S. acknowledges the support from the Swedish Research Council (VR) through a Starting Grant (Dnr. 2017-05078).

# Figures

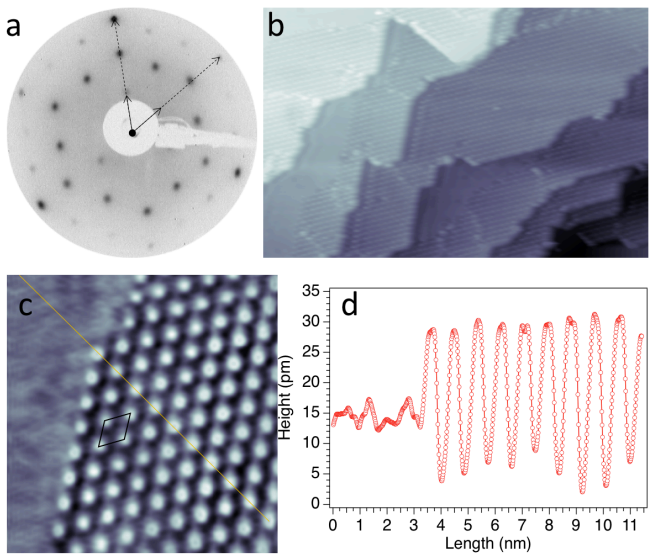

Fig. 1: **Atomic arrangement of the Al(111)3×3-Si surface, as prepared directly at room temperature**. a) LEED pattern (E0 = 50 eV). b) 44 nm × 29 nm topograph (IT = 170 pA, VGAP = 80 mV, empty states). c) 9 nm × 9 nm topograph (IT = 170 pA, VGAP = -100 mV, filled states), showing a Si covered 3×3 island (right part) besides a bare Al(111) region (left part). d) line profile of the 3×3 reconstruction of 1 ML of Si on Al(111). Note that the size of the lozenge in (a) is 8.6 Å, i.e., 3aAl(111) (with aAl(111) = 2.86 Å, the in-plane lattice parameter of the bare Al(111) surface), in accord with the 3×3 superstructure observed in the LEED patterns, as in a).

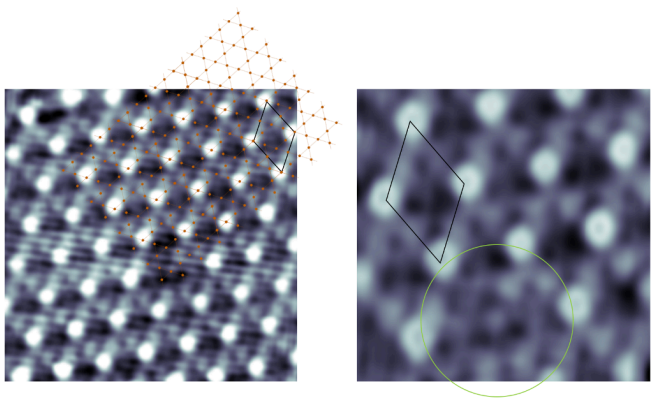

Fig. 2: **Kagome-like lattice of the Al(111)3×3-Si surface.** 6 nm × 6 nm (left) and 3 nm × 3 nm (right) high-resolution STM images (IT = 320pA, VGAP = 80 mV, empty states) of the Al(111)3×3-Si surface. For the sake of the discussion, a 2D

Kagome lattice (orange mesh) has been superposed on one part on the STM image in the left panel. The observed superstructure adopts a 3×3 surface reconstruction with respect to the Al(111), corresponding to a (√3×√3)R30° reconstruction with respect to a pristine 2D Kagome lattice.

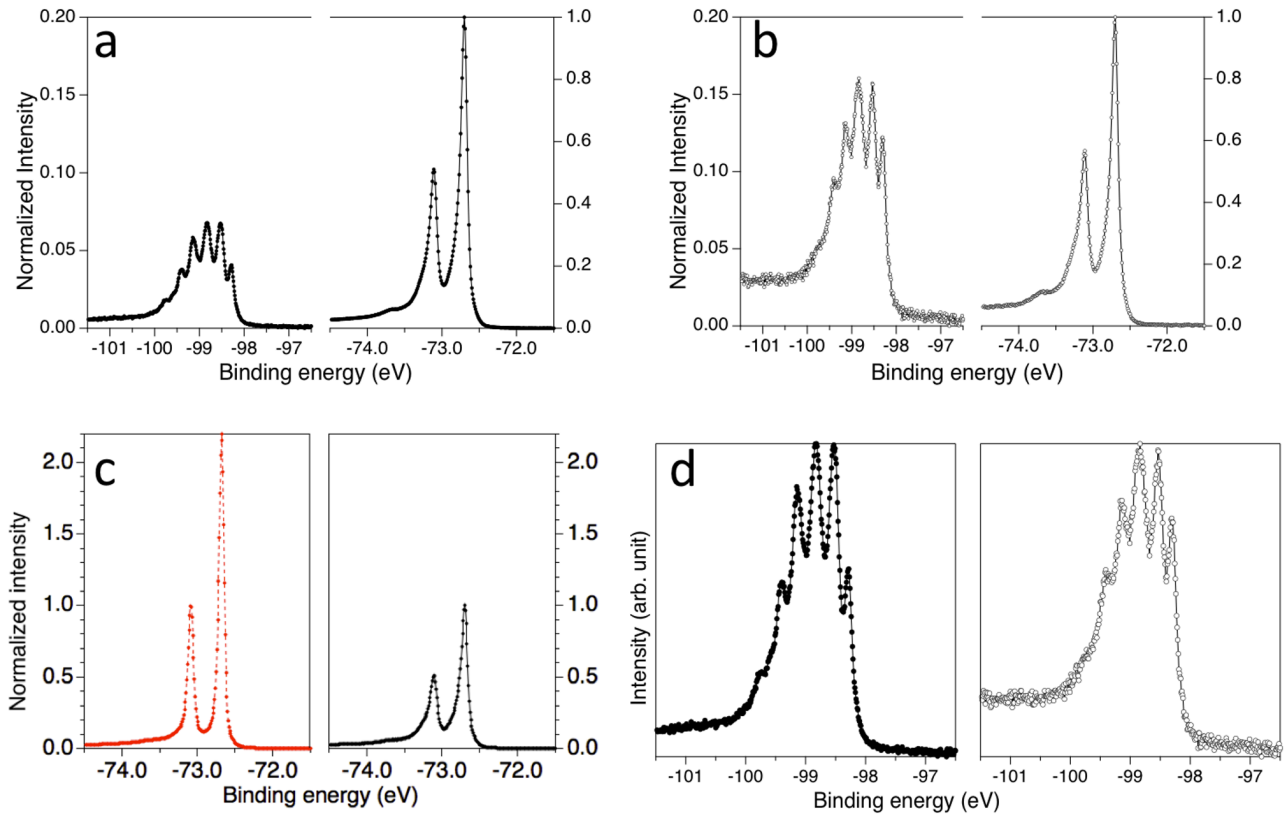

Fig. 3: **Al 2p and Si 2p core level photoelectron spectra taken at 136 eV photon energy for the Al(111)3×3-Si surface**. The full markers correspond to data recorded in normal emission, the hollow markers correspond to data recorded 60° off-normal. The spectrum in red in panel c was acquired on the clean Al(111) surface. Spectra in panels a) and b) have been normalized to the Al 2p3/2 line.

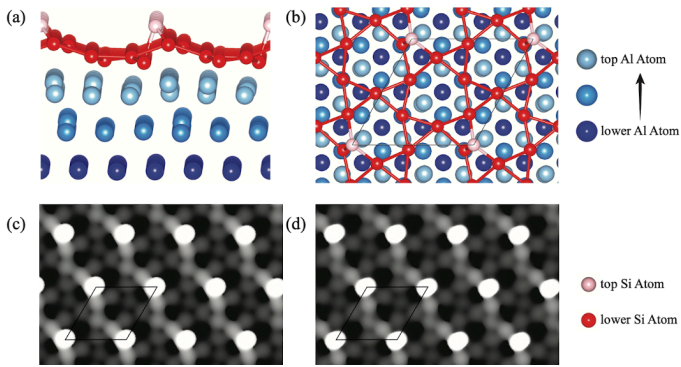

Fig. 4: **Kagome silicene with one DB per Al(111)-3×3 supercell.** Atomic structure models in a) side and b) top views. c-d) Simulated STM images (bias = - 0.1 / + 0.1 V) of filled (c) and empty (d) states.

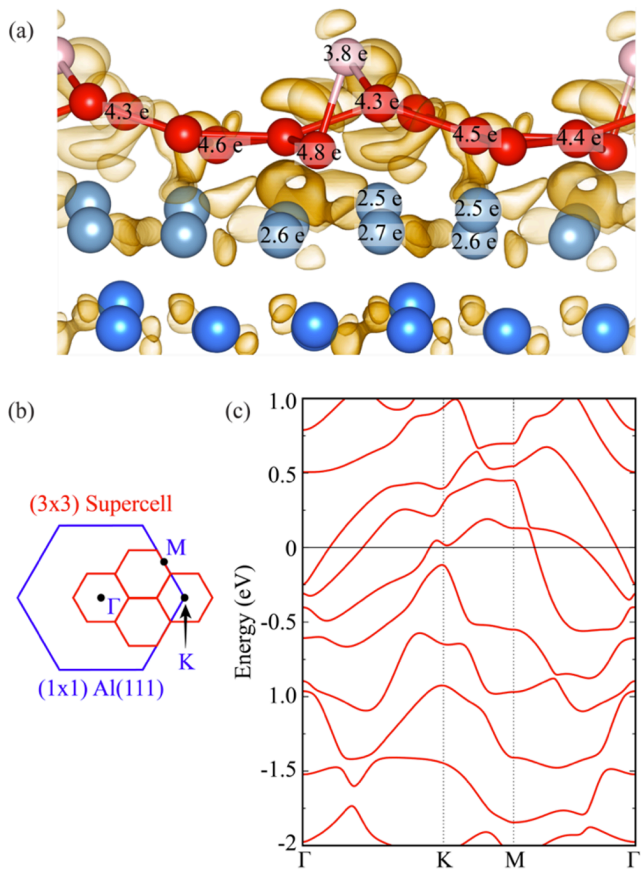

Fig. 5: **Interatomic charge transfers calculation**. a) The charge density map of the whole structure. b) The charge densities of the individual Al and Si atoms subtracted from the charge density of the whole structure; valence charges are also shown on the atoms. c) Scheme of the Brillouin Zones of 1x1 Al(111) (in blue) and the 3×3 superstructure (in red). d) The band structure of the freestanding Kagome silicene layer.

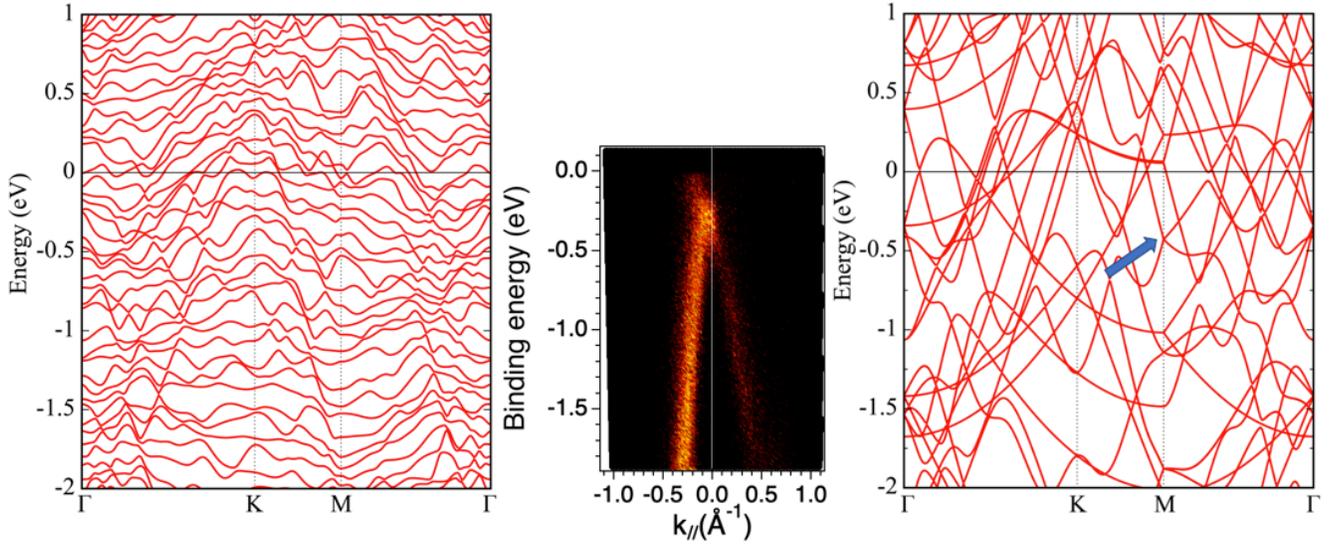

Fig. 6: **Nearly linear dispersions along $\overline{\Gamma}\overline{M}$**. Left-panel) Electronic band structure of Kagome Silicene on Al(111)-3×3. Middle-panel) ARPES data record with HeI on the Al(111)3×3-Si surface around the $\overline{M}$ point of the 3×3 superstructure, along the $\overline{\Gamma}\overline{M}$ direction. Right-panel) Folded 3×3 band structure calculated for a clean Al(111).